Enhanced ionic conductivity through crystallization of glass-$Li_3PS_4$ by machine learning molecular dynamics simulations

Koji Shimizu [a]*, Parth Bahuguna [b], Shigeo Mori [c], Akitoshi Hayashi [d], and Satoshi Watanabe [a]*

[a] Department of Materials Engineering, The University of Tokyo, 7-3-1 Hongo, Bunkyo-ku, Tokyo 113-8656, Japan

[b] Department of Physics, Indian Institute of Technology Roorkee, Roorkee 247667, India

[c] Department of Material Science, Osaka Metropolitan University, 1-1 Gakuen-cho, Naka-ku, Sakai, Osaka 599-8531, Japan

[d] Department of Applied Chemistry, Osaka Metropolitan University, 1-1 Gakuen-cho, Naka-ku, Sakai, Osaka 599-8531, Japan

**Abstract**: Understanding the atomistic mechanism of ion conduction in solid electrolytes is critical for the advancement of all-solid-state batteries. Glass-ceramics, which undergo crystallization from a glass state, frequently exhibit unique properties including enhanced ionic conductivities compared to both the original crystalline and glass forms. Despite these distinctive features, specific details regarding the behavior of ion conduction in glass-ceramics, particularly concerning conduction pathways, remain elusive. In this study, we demonstrate the crystallization process of glass-$Li_3PS_4$ through molecular dynamics simulations employing machine learning interatomic potentials constructed from first principles calculation data. Our analyses of Li conduction using the obtained partially crystallized structures reveal that the diffusion barriers of Li decrease as the crystallinity in $Li_3PS_4$ glass-ceramics increases. Furthermore, Li displacements predominantly occur in the precipitated crystalline portion, suggesting that percolation conduction plays a significant role in enhanced Li conduction. These findings provide valuable insights for the future utilization of glass-ceramic materials.

Exploring fast ion-conducting solid electrolytes is essential to achieve the practical application of all-solid-state lithium (Li) batteries, which is anticipated for their high safety and energy density. Among the various solid electrolyte materials developed thus far, $Li_{10}GeP_2S_{12}$ (LGPS), a sulfide solid electrolyte, exhibits an ionic conductivity of 12 mS cm$^{-1}$ at room temperature [1], which is comparable to that of liquid electrolytes. Subsequently, $Li_{9.54}Si_{1.74}P_{1.44}S_{11.7}Cl_{0.3}$ [2], a member of the LGPS family, and $Li_{6.6}Ge_{0.6}P_{0.4}S_5I$ [3], an argyrodite-type material, were identified to exhibit even higher ionic conductivities. While several challenges, including the material stability and compatibility with electrode materials, remain to be addressed, sulfur-based solid electrolytes emerge as promising candidates, which highlights the significance of gaining a more detailed understanding of these materials.

The mixtures of $Li_2S$-$P_2S_5$ have been extensively studied experimentally owing to their enhanced ionic conductivity observed in the glass phases [4, 5, 6]. Furthermore, studies have reported that ionic conductivity further increases when glass structures crystallize through heat treatment, resulting in the formation of glass-ceramics [7, 8]. Regarding glass-$Li_3PS_4$ (a mixture of 75$Li_2S$·25$P_2S_5$ in mol%), recent experiments revealed the stabilization of the high-temperature crystalline phase ($\alpha$-$Li_3PS_4$) at room temperature through rapid heating [9]. These findings suggest that glass-ceramics exhibit unique characteristics and the potential to exploit the physical properties of phases that are not thermodynamically stable at room temperature.

In atomistic simulations, density functional theory (DFT) calculations serve as a powerful tool for analyzing phenomena in complex systems. Previously, using ab initio molecular dynamics (AIMD) simulations, the ion-conducting mechanism in glass-$Li_3PS_4$ was thoroughly examined [10]. However, an understanding of the ionic behavior at the phase boundary between the glass and crystalline regions, which is distinctive in glass-ceramics, remains limited. Modeling of such glass-ceramics structures naturally necessitates large systems, and accordingly, substantial computational costs prevent addressing these challenges.

Considering these circumstances, machine learning (ML) of interatomic potentials has gained increasing attention. Starting with pioneering methods, such as the high-dimensional neural network potential (NNP) [11] and Gaussian approximation potential [12], various new approaches, including graph-NN based models [13, 14, 15], have been progressively developed. A few of these models have been applied to solid electrolyte materials, demonstrating a comparable prediction accuracy of defect and ion-conducting properties to DFT calculations with low computational costs [16, 17]. Additionally, an NNP study of $x$$Li_2S$·(1-$x$)$P_2S_5$ has demonstrated its capability to reveal thermodynamically stable motifs across a wide compositional space of $x$ [18].

Based on the aforementioned, in this study, we constructed an NNP for $Li_3PS_4$ using DFT calculation data. Subsequently, a glass structure was generated through dynamical simulations, followed by its crystallization via an annealing procedure. Further analysis was conducted on the ion-conducting behavior in the partially crystallized structures obtained. The calculation results provided microscopic insights into the phase changes and ionic behavior in glass-ceramics.

**Results and discussion**

**Construction of machine learning potentials**

We constructed an NNP using a structural dataset generated through AIMD simulations under various conditions. Among the extracted 34,510 structural data from the AIMD trajectories, 80% (20% remaining) of the data was randomly chosen as the training (test) data, and the resultant root-mean-square errors (RMSEs) of the total energies and atomic forces were 12.5 (12.7) meV/atom and 240 (241) meV/Å, respectively. The parity plots between the DFT reference values and the NNP predictions are provided in Supplementary Fig. 1. While the constructed NNP can predict all the data points with a reasonably high accuracy, the obtained RMSE values remained slightly large compared to those reported in other studies using ML potentials. AIMD simulations of $Li_3PS_4$ with temperatures higher than 2000 K resulted in strongly distorted structures, including broken $PS_4^{3-}$ units. In addition to the structural complication, this led to a large modulation of their electronic states; in fact, the RMSEs of trainings using structural data below 2000 K reached less than 10 meV/atom and 100 meV/Å for the total energies and atomic forces, respectively. To obtain a significantly higher accuracy, sophisticated architectures other than symmetry functions are necessary to accurately capture the atomic environments, which will be addressed by using the aforementioned graph-NN-based methods in the future.

**Vitrification of crystalline-$Li_3PS_4$**

To generate a glass-$Li_3PS_4$ structure, we performed canonical ensemble (*NVT*) molecular dynamics (MD) simulations by employing a melt-quench method. Figure 1a depicts a schematic of the initial crystalline (β-phase) structure, which consisted of 5760 atoms/supercell. Initially, the temperature was raised from 300 to 1500 K at a rate of 240 K/ps over 5 ps, which was maintained at 1500 K for 10 ps. Subsequently, the system underwent cooling at a rate of 4 K/ps over 300 ps, followed by equilibration at 300 K for 7.5 ps. Changes in the temperature and potential energy during the melt-quench calculations are depicted in Fig. 1b.

Figure 1c illustrates a schematic of the resultant $Li_3PS_4$ structure after the equilibration step. The calculated pair distribution functions (PDFs) of the corresponding structure are depicted in Fig. 1d. The total PDF indicated no specific features beyond a distance of 4 Å, indicating the glass nature of the obtained structure. In addition, the peak positions and shapes in the PDF profiles closely match those reported by a previous study utilizing AIMD simulations [10] for all elemental combinations. This agreement highlights the high accuracy of the present NNP.

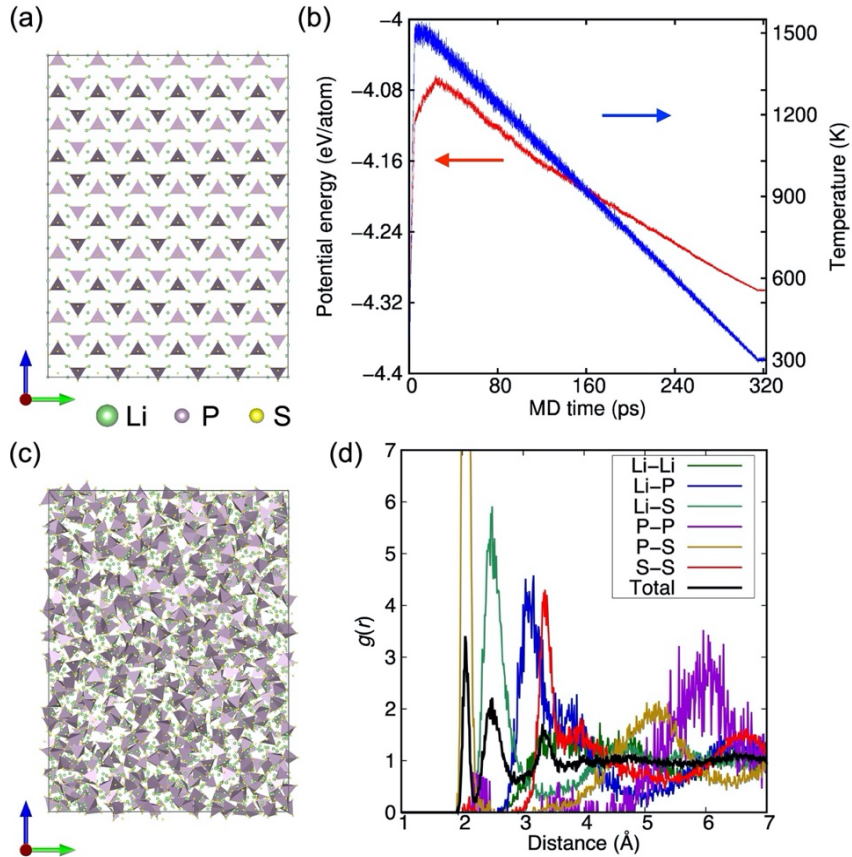

Fig. 1 **The vitrification process of Li$_3$PS$_4$.** (a) Schematic of the β-Li$_3$PS$_4$ structure. (b) Potential energy (red) and temperature (blue) profiles during the MD simulations. (c) Schematic of glass-Li$_3$PS$_4$ obtained by the melt-quench process. (d) Calculated pair distribution functions of the corresponding glass-Li$_3$PS$_4$. The atomic structures were visualized by the VESTA software [37].

**Crystallization of glass-Li$_3$PS$_4$**

Glass-ceramics are generally produced by heating glass materials in experiments. To examine the formation of glass-ceramics through heat treatment, we performed annealing MD simulations in the isothermal-isobaric ensemble (*NPT*) at a pressure of 1 bar and temperature of 600 K, starting from the glass-Li$_3$PS$_4$ structure obtained through the melt-quench process. Figure 2a depicts the calculated potential energy profile as a function of the MD time. Notably, the potential energy remained nearly constant until approximately 70 ns, after which a substantial decrease occurred between 70 to 85 ns, suggesting that the system underwent a phase change during that period.

To assess changes in the crystallinity owing to the decrease in the potential energy, we performed a series of X-ray diffraction (XRD) analyses on the annealed structures, ranging from 70 to 100 ns with 1 ns intervals, subsequent to structural optimization. The resultant crystallinity values, derived from the XRD patterns, are presented in Fig. 2b. At the annealing time of 70 ns, prior to the decrease in potential energy, the crystallinity was notably low (< 10%), providing another indicator of the glass state. Subsequently, after 70 ns, the crystallinity gradually increased to approximately 50%. As the crystallinity increased, the cell volume gradually decreased, as depicted in Fig. 2b. These crystallization results were evident in the XRD profiles presented in Fig. 2c. Prior to annealing (0 ns), no significant visible peaks were observed, whereas the peak intensities started to increase after the annealing time of 80 ns.

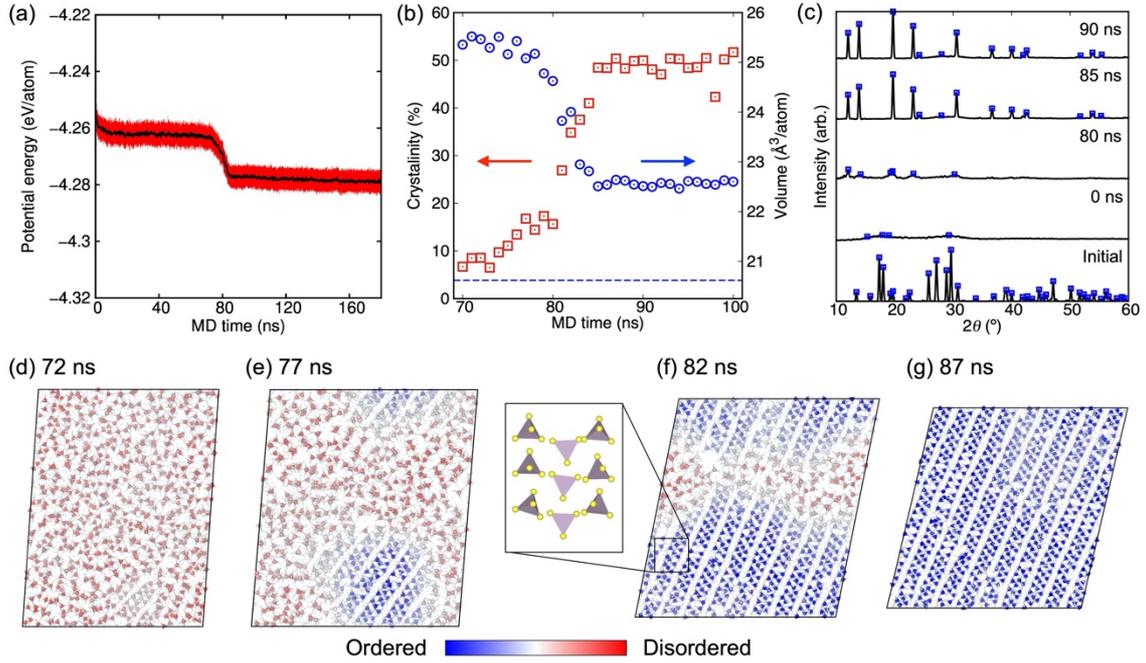

Fig. 2 **Crystallization of glass-Li$_3$PS$_4$.** (**a**) Calculated potential energy profile as a function of the MD time; the red and black lines indicate the overall and averaged values, respectively. (**b**) Changes in the crystallinity (red) and the cell volume (blue) by the decrease in potential energy. Prior to these evaluations, structural optimizations were conducted at specific annealing times. (**c**) Calculated XRD patterns of the initial (β-phase) and glass-Li$_3$PS$_4$ (0 ns) structures, and the structures annealed over 80, 85, and 90 ns. Schematics of the resultant structures after annealing over (**d**) 72, (**e**) 77, (**f**) 82, and (**g**) 87 ns. The colors of atoms indicate the degree of structural ordering, wherein blue and red indicate ordered and disordered structures, respectively. The inset in (**f**) illustrates a pattern of the PS$_4$ arrangement at the crystallized region. The atomic structures were visualized by the OVITO software [38].

Figure 2d-g depicts the snapshot structures at annealing times of 72, 77, 82, and 87 ns, wherein the colors of the atoms indicate the degree of ordering of the phosphorus atoms (Li and S atoms are depicted in white), as estimated by the entropy-based fingerprint (EBF) method [19]. Notably, crystalline nucleation initiated in the right-bottom region and gradually expanded as the annealing proceeded. At 77 ns, the crystalline nucleus was surrounded by glass regions. In contrast, by 82 ns,

the crystalline nucleus grew and connected both ends in the lateral direction. Furthermore, by 87 ns, the crystalline nucleus further expanded, connecting the other ends in the longitudinal direction, and covering the entire region. The crystallinity reached approximately 50% at this point and remained nearly constant until 100 ns. Locally disordered portions within the structures hindered further progress in crystallinity. Note, the crystalline nucleus was not perfectly isolated from the glass regions, as it demonstrated periodicity along the direction perpendicular to the figure from the beginning.

The crystalline phases ($\alpha$, $\beta$, and $\gamma$) of $Li_3PS_4$ are characterized by the arrangement of the $PS_4$ units. The inset in Fig. 2f depicts a part of the $PS_4$ unit arrangement in the crystallized region, which aligns with the specific features of the $\alpha$-$Li_3PS_4$ phase. Additionally, the XRD profiles of the crystallized structures exhibit a single peak at $2\theta \sim 20°$, which is consistent with the experimental profile of the $\alpha$-phase [9]. Note, a slightly larger angle compared to the experimental and our DFT calculation data ($2\theta \sim 17\text{-}19°$) may indicate a need for improving the accuracy of our potential. Furthermore, recent experiments revealed that rapidly heating glass-$Li_3PS_4$ up to 550 K results in its precipitation into the $\alpha$-phase [9], and DFT calculations demonstrated the stability of $\alpha$-phase above 460 K [20]. These experimental and theoretical findings once again align with our resultant structures annealed at 600 K.

**Li conductivity calculations**

As described in the previous section, we successfully achieved the crystallization of glass-$Li_3PS_4$. The conductivities of Li based on the degree of crystallinity were subsequently investigated by utilizing the structures at different annealing times. Figure 3a, b depicts the calculated diffusion coefficients of Li at 400 and 600 K, respectively. The simulations utilized the structures obtained through annealing times ranging from 70 to 100 ns at 1 ns intervals, subsequent to structural optimization (see Supplementary Fig. 2 for schematics of the three-dimensional structures). Additionally, the diffusion coefficient values for the initial crystalline ($\beta$-phase) and glass structures are provided. It is noteworthy that the simulation results sufficiently captured the enhanced Li conductivity resulting from the vitrification of $\beta$-$Li_3PS_4$.

At 400 K, the diffusion coefficients of Li increased as the crystallinity increased. Conversely, an opposite trend was observed at 600 K. These results suggest that the diffusion behaviors of Li in the partially crystallized structures differ from both their initial crystalline and glass states. Based on the Arrhenius plots shown in Fig. 3c, we evaluated the diffusion barrier heights of Li, as depicted in Fig. 3d. The glass-$Li_3PS_4$ exhibited a lower diffusion barrier of 0.269 eV than that of the initial crystalline ($\beta$-phase) with a value of 0.311 eV. A previous study reported the diffusion barrier of $\beta$-$Li_3PS_4$ as approximately 0.3 eV, which was obtained by the nudge elastic band calculations based on DFT [21]. Additionally, the diffusion barrier of glass-$Li_3PS_4$ was estimated as 0.22 or 0.25 eV in previous AIMD simulations [10]. These reported values sufficiently agreed with the present results. Furthermore, the Li conductivity in the glass-Li$_3$PS$_4$ was estimated as 7.45×10$^{-3}$ S cm$^{-1}$ at 300 K based on the Nernst-Einstein equation. Once again, this was in fairly good agreement with previous AIMD simulations, which reported values of 1.9×10$^{-2}$ or 7.0×10$^{-3}$ S cm$^{-1}$ [10]. Note, in this study, the diffusion barrier of Li in the glass-Li$_3$PS$_4$ was slightly higher than that of the AIMD simulations, which may be attributed to the fact that the present glass-Li$_3$PS$_4$ structure included complex anion units, that is, P$_2$S$_6^{4-}$ and P$_2$S$_7^{4-}$, as well as Li-S and S-S species (see Supplementary Fig. 3). This can be rationalized mainly by the higher melting temperature and additionally by the larger model size, which accommodated the higher structural degree of freedom. Note, the calculated Li conductivity remained to overestimate the experimental values [4, 5, 6]. The experiments generated glass structures through the mechanical milling method, which may have resulted in structures with different features from those fabricated by melt-quenching. Nonetheless, the melt-quenching process is currently the only plausible method of generating glass structures in atomistic simulations.

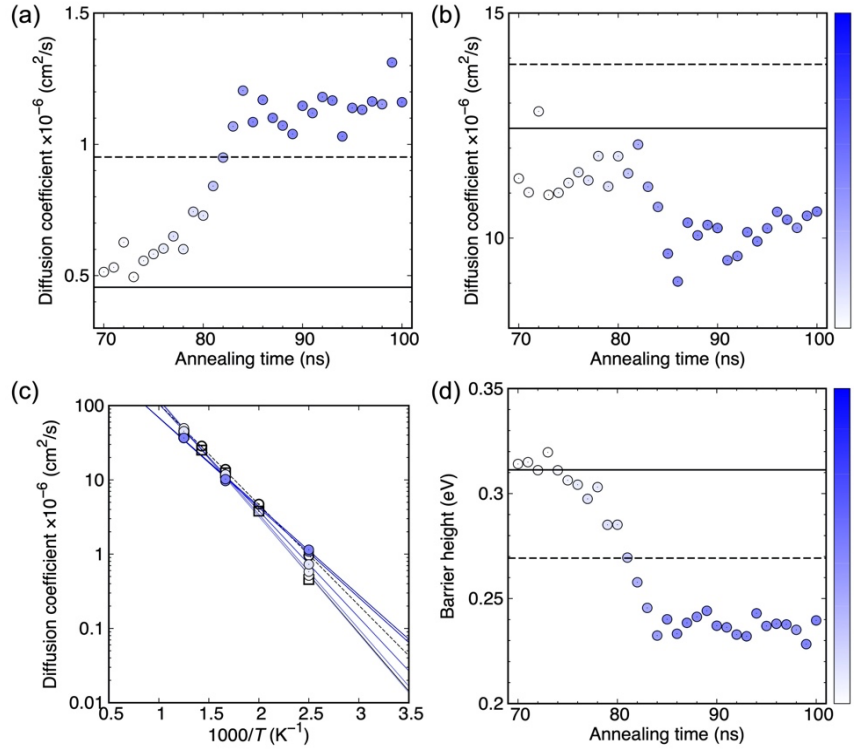

Fig. 3 **Diffusion coefficients and barrier heights of Li.** Calculated diffusion coefficients of Li at **(a)** 400 K and **(b)** 600 K. The optimized structures obtained after annealing at specific times were employed for these simulations. The colored symbols indicate the corresponding crystallinity values. The solid and dotted lines indicate the values for the β- and glass-Li$_3$PS$_4$ structures, respectively. **(c)** Diffusion coefficients as a function of the inverse temperature for structures obtained after annealing for 70, 75, 80, 85, and 90 ns. **(d)** Barrier heights of the Li diffusion evaluated from the Arrhenius relationship.

At the annealing time of 70 ns, the diffusion barrier of Li became higher and approached that of the initial crystalline (β-Li$_3$PS$_4$) structure. The increase in the barrier height suggested that structural orderings gradually proceeded to a certain extent, despite the absence of crystalline nucleation. This ordering presumably hindered Li diffusion in the glass regions. After crystalline nucleation, the diffusion barrier of Li gradually decreased as the crystallinity increased. The partially crystallized structure that was annealed for longer than 80 ns exhibited a lower diffusion barrier of Li than that of the glass structure. At 84 ns, it reached a nearly constant value of 0.232 eV as the crystallization expanded to all the regions.

Subsequently, to analyze the regions and pathways where Li frequently moved in the partially crystallized structures, we evaluated the Li displacements using models with different annealing times. Figure 4a-d depicts snapshots of the glass structure and structures annealed for 70, 80, and 84 ns, along with coloring by the EBF method. The amplitudes of the Li displacements in the corresponding structures are illustrated in Fig. 4e-h through the color mappings. Regarding glass-Li$_3$PS$_4$ (Fig. 4e), Li moved almost uniformly, with higher amplitudes throughout the entire regions, as indicated by the red colors. The aforementioned complex anion units, that is, P$_2$S$_6^{4-}$ and P$_2$S$_7^{4-}$, seemed to impede Li diffusion in the blue-colored regions. In comparison, the Li displacements were largely suppressed in the structure annealed over 70 ns (Fig. 4f) prior to crystallization, despite the P$_2$S$_7^{4-}$, Li-S, and S-S species being mostly rearranged to reform PS$_4^{3-}$ (see Supplementary Fig. 3). Additionally, regions with higher Li displacements were scattered. In this structure, non-uniformity in the local Li density was observed among the glass regions (see Supplementary Fig. 4), which potentially acted as a barrier to hinder the Li diffusion paths. At 80 ns (Fig. 4g), in turn, the amplitude of the Li displacements increased, which were notably clustered. The clustered regions clearly coincided with the crystallized portion. Furthermore, as the crystallization proceeded, regions with higher Li displacements demonstrated grid lines, which indicated Li motion along the paths existing in the crystalline structure.

Based on the aforementioned results, the reduction in the diffusion barrier can be attributed to the presence of precipitated crystalline components, which were identified as the α-phase, known for its characteristics as a high-temperature phase and fast ion-conductor. It is noteworthy that the present NNP predicted the diffusion barrier of α-Li$_3$PS$_4$ as 0.198 eV, which closely aligns with the value of 0.18 eV obtained through AIMD simulations [22]. Furthermore, the diffusion barrier rapidly decreased from 80 to 84 ns of the annealing time (see Fig. 3d). The crystalline nucleus gradually grew, and laterally connected within the cell at 81 ns. Additionally, the growth of the crystalline nucleus horizontally extended until 83 ns. These results suggest that additional Li diffusion pathways were formed by contacting the crystalline nuclei. Thus, we concluded that the significant factor contributing to the enhanced Li conductivity in Li$_3$PS$_4$ glass-ceramics is not the interface between the crystalline and glass structures, but rather the percolation conduction. Note, the predominant ion-conducting

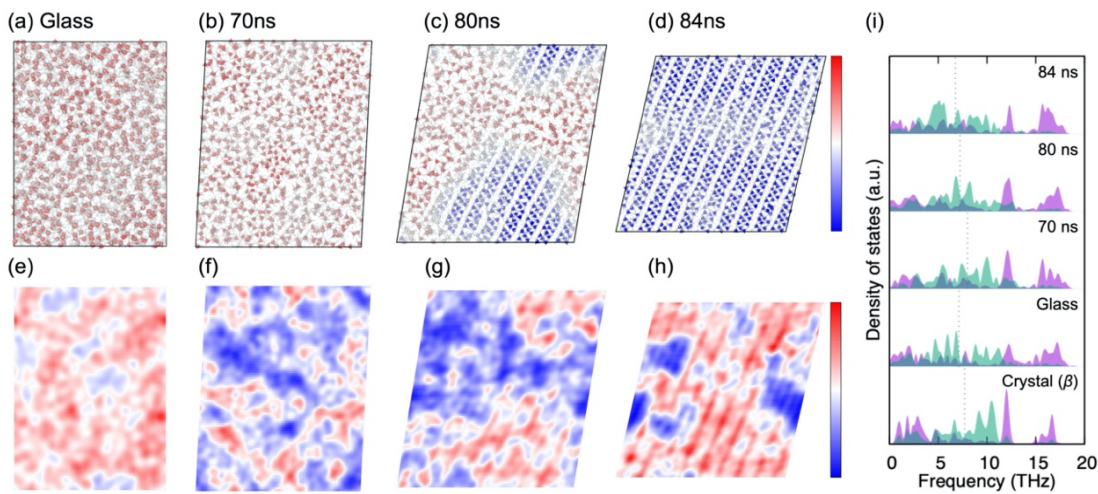

Fig. 4 **Mappings of Li displacements and vibrational density of states.** Schematics of **(a)** the glass-$Li_3PS_4$ and the structures obtained after annealing for **(b)** 70, **(c)** 80, and **(d)** 84 ns. The colors of atoms indicate the degree of structural ordering, wherein blue and red indicate ordered and disordered structures, respectively. Mappings of Li displacements at **(e)** the glass-$Li_3PS_4$ and annealed structures at **(f)** 70, **(g)** 80, and **(h)** 84 ns. Red (blue) indicates regions with larger (smaller) Li displacements. **(i)** Calculated vibrational density of states (VDOS) of β- and glass-$Li_3PS_4$, and the structures obtained through annealing for 70, 80, and 84 ns. The green and purple colors indicate the contributions from cations (Li) and anions (P and S), respectively. The dotted lines depict the band centers of Li VDOSs.

mechanism may vary depending on the precipitated crystalline phase [23].

Finally, we calculated the vibrational density of states (VDOS) of the ions from the Fourier transformation of the velocity-velocity autocorrelation functions. Figure 4i depicts the VDOSs of the corresponding structures, including the initial crystalline (β-phase) structure, wherein the contributions from anions (P and S) and cations (Li) are separately provided. The center position of the Li VDOS in the glass structure is located at a lower frequency compared to that of the initial crystalline (β-phase) structure, which agrees with a previously reported indicator [24], wherein Li conductivities correlated with its VDOS centers. Lowering of the center position suggests that Li undergoes softer potential energy surfaces, leading to larger displacement amplitudes of Li. Furthermore, the VDOS center of Li in the structure annealed over 70 ns shifted toward a higher frequency. Li ions in regions of lower density, as depicted in Supplementary Fig. 4, experienced deeper potentials, leading to an increase in the vibrational frequency. In comparison, as annealing proceeded, the VDOS centers shifted toward lower frequencies, which was consistent with the change in diffusion barriers during crystallization, even in complex systems.

In summary, we investigated the crystallization process of glass-$Li_3PS_4$ through MD simulations employing ML potentials constructed from DFT calculation data. The calculated diffusion barriers of Li in the obtained partially crystallized structures demonstrated lower values as the crystallinity in Li$_3$PS$_4$ glass-ceramics increased, whereas the barriers increased with heating prior to the occurrence of crystallization. Furthermore, Li displacements predominantly occurred at the precipitated crystalline portion, which was confirmed as the high-temperature α-phase. The diffusion barriers of Li rapidly decreased when the grown crystalline nuclei made contact in the lateral and horizontal directions, suggesting that the enhanced Li conduction can be attributed to percolation conduction rather than conduction at the interface of the crystalline and glass region. These findings provide valuable insights for the future utilization of glass-ceramic materials.

## Methods

### Density functional theory calculations

To generate a structural dataset for training a NNP, we first performed AIMD simulations. Two types of crystalline $Li_3PS_4$ structures obtained from the Materials Project database [25, 26] (mp-985583 and mp-1097036) [19, 20] were employed as the initial structures. We used models containing 32 atoms/supercell and 64 atoms/supercell (enlarged by two times along the *a*-axis) with the optimal lattice constant (see Supplementary Fig. 5), as well as that compressed and expanded by ±2%. In addition to the pristine models, we also considered defective structures by introducing one Li or $Li_2S$ vacancy into the supercells. The AIMD simulations were performed in the canonical ensemble (*NVT*) at 500, 1000, 1500, 2000, 3000, and 4000 K for 100 ps, with a timestep of 2 fs (50000 steps) in each condition. Subsequently, we uniformly extracted 5% of snapshots from the trajectories. Furthermore, employing the methodology of our previous study [27], we screened structures based on similarities evaluated by symmetry functions (SFs), and finally obtained 34,510 structures, for which static DFT calculations were performed under severe conditions.

We used the generalized gradient approximation with the Perdew-Burke-Ernzerhof functional [28], the planewave basis set (400 and 600 eV cutoff energies for the AIMD and static DFT calculations, respectively), and projector-augmented wave method [29]. Brillouin zone integration was performed using the sampling technique of Monkhorst and Pack [30]; 2×2×1 (4×4×2) and 1×2×1 (2×4×2) for 32 and 64 atom models, respectively, in the AIMD (static DFT) calculations. The self-consistent calculation for the electronic states was conducted with the convergence criterion of $10^{-4}$ ($10^{-7}$) eV in the AIMD (static DFT) calculations. We used the *Vienna Ab initio Simulation Package* software [31, 32] for all the DFT calculations.

### Neural network potential trainings

As an ML potential of $Li_3PS_4$, we employed the Behler-Parrinnelo-type NNP model [11]. The radial and angular SFs were used to describe the local atomic features with a cutoff distance of 7 Å. In each elemental combination, we used 7 and 24 different parameters for radial and angular SFs, respectively (see Supplementary Note 1 for details regarding SFs). We used the NN consisting of two hidden layers with 10 nodes and a single output layer. Thus, the NNP architecture was [165-10-10-1]. We trained the NNP using the loss of total energies and atomic forces obtained by the DFT calculations.

### Molecular dynamics simulations

To investigate the structural changes and Li diffusion behaviors, we performed MD simulations using models containing 5760 atoms/supercell. The initial crystalline (β-phase) structure was prepared by enlarging the MP-985583 by 6×6×5. For the Li conductivity calculations, we performed *NVT*-MD simulations for 1 ns, wherein the diffusion coefficients of Li were evaluated by the linear fitting of mean-square-displacements from 0.1 to 1 ns. For mapping the Li displacements, we assessed the amplitude of the Li displacements every 1 ps and recorded these values at the corresponding Li positions before the displacements; this process was continued for 1 ns. A timestep of 1 fs was used for all these calculations. The *LAMMPS* software [33] was used in this study.

The XRD profiles were generated using the *pyxtal* package [34]. The crystallinity was estimated from the ratio of the integrated peak area in the XRD profile to that of the entire region. Peaks were identified using the *find_peaks* module in the *SciPy* package [35]. The EBF method was applied to phosphorus atoms with parameters of $r_m$ = 10 Å, $r_a$ = 9 Å, and $\sigma$ = 0.1 Å. The VDOSs were computed using the *dump2vdos* code [36], utilizing the velocity-velocity autocorrelation functions obtained from the *NVT*-MD simulations conducted for 1 ns at 100 K. The VDOS centers were determined by averaging the frequency from the numerical integrations.

**Acknowledgements**.

We thank Mr. Yusei Tomenaga and Mr. Clemen Goh for their contributions in the early stage of this study, which was supported by JSPS KAKENHI Grant Numbers 20H05285, 22H04607, and 23H04100, and Editage (www.editage.com) for English language editing. Certain calculations used in this study were performed using the computer facilities at the ISSP Supercomputer Center and Information Technology Center, The University of Tokyo, and Institute for Materials Research, Tohoku University.

References

[1] N. Kamaya, K. Homma, Y. Yamakawa, M. Hirayama, R. Kanno, M. Yonemura, T. Kamiyama, Y. Kato, S. Hama, K. Kawamoto and A. Mitsui, A lithium superionic conductor, vol. 10, Nat. Mater., 2011, p. 682.

[2] Y. Kato, S. Hori, T. Saito, K. Suzuki, M. Hirayama, A. Mitsui, M. Yonemura, H. Iba and R. Kanno, High-power all-solid-state batteries using sulfide superionic conductors, vol. 1, Nat. Energy, 2016, p. 16030.

[3] M. A. Kraft, S. Ohno, T. Zinkevich, R. Koerver, S. P. Culver, T. Fuchs, A. Senyshyn, S. Indris, B. J. Morgan and W. G. Zeier, Inducing High Ionic Conductivity in the Lithium Superionic Argyrodites $Li_{6+x}P_{1-x}Ge_xS_5I$ for All-Solid-State Batteries, vol. 140, J. Am. Chem. Soc., 2018, p. 16330.

[4] A. Hayashi, S. Hama, H. Morimoto, M. Tatsumisago and T. Minami, Preparation of $Li_2S$–$P_2S_5$ Amorphous Solid Electrolytes by Mechanical Milling, vol. 84, J. Am. Ceram. Soc., 2001, p. 477.

[5] K. Ohara, A. Mitsui, M. Mori, Y. Onodera, S. Shiotani, Y. Koyama, Y. Orikasa, M. Murakami, K. Shimoda, K. Mori, T. Fukunaga, H. Arai, Y. Uchimoto and Z. Ogumi, Structural and electronic features of binary $Li_2S$-$P_2S_5$ glasses, vol. 6, Sci. Rep., 2016, p. 21302.

[6] C. Dietrich, D. A. Weber, S. J. Sedlmaier, S. Indris, S. P. Culver, D. Walter, J. Janek and W. G. Zeier, Lithium ion conductivity in $Li_2S$–$P_2S_5$ glasses – building units and local structure evolution during the crystallization of superionic conductors $Li_3PS_4$, $Li_7P_3S_{11}$ and $Li_4P_2S_7$, vol. 5, J. Mater. Chem. A, 2017, p. 18111.

[7] F. Mizuno, A. Hayashi, K. Tadanaga and M. Tatsumisago, New, Highly Ion-Conductive Crystals Precipitated from $Li_2S$-$P_2S_5$ Glasses, vol. 17, Adv. Mater., 2005, p. 918.

[8] H. Tsukasaki, S. Mori, S. Shiotani and H. Yamamura, Ionic conductivity and crystallization process in the $Li_2S$–$P_2S_5$ glass electrolyte, vol. 317, Solid State Ionics, 2018, p. 122.

[9] T. Kimura, T. Inaoka, R. Izawa, T. Nakano, C. Hotehama, A. Sakuda, M. Tatsumisago and A. Hayashi, Stabilizing High-Temperature a-$Li_3PS_4$ by Rapidly Heating the Glass, vol. 145, J. Am. Chem. Soc., 2023, p. 14466.

[10] J. G. Smith and D. J. Siegel, Low-temperature paddlewheel effect in glassy solid electrolytes, vol. 11, Nat. Commun., 2020, p. 1483.

[11] J. Behler and M. Parrinello, Generalized Neural-Network Representation of High-Dimensional Potential-Energy Surfaces, vol. 98, Phys. Rev. Lett., 2007, p. 146401.

[12] A. P. Bartók, M. C. Payne, R. Kondor and G. Csányi, Gaussian Approximation Potentials: The Accuracy of Quantum Mechanics, without the Electrons, vol. 104, Phys. Rev. Lett., 2010, p. 136403.

[13] K. T. Schütt, H. E. Sauceda, P. -J. Kindermans, A. Tkatchenko and K. -R. Müller, SchNet – A deep learning architecture for molecules and materials, vol. 148, J. Chem. Phys., 2018, p. 241722.

[14] S. Batzner, A. Musaelian, L. Sun, M. Geiger, J. P. Mailoa, M. Kornbluth, N. Molinari, T. E. Smidt and B. Kozinsky, E(3)-equivariant graph neural networks for data-efficient and accurate interatomic potentials, vol. 13, Nat. Commun., 2022, p. 2453.

[15] A. Musaelian, S. Batzner, A. Johansson, L. Sun, C. J. Owen, M. Kornbluth and B. Kozinsky, Learning local equivariant representations for large-scale atomistic dynamics, vol. 14, Nat. Commun., 2023, p. 579.

[16] W. Li, Y. Ando, E. Minamitani and S. Watanabe, Study of Li atom diffusion in amorphous $Li_3PO_4$ with neural network potential, vol. 147, J. Chem. Phys., 2017, p. 214106.

[17] A. Marcolongo, T. Binninger, F. Zipoli and T. Laino, Simulating Diffusion Properties of Solid-State Electrolytes via a Neural Network Potential: Performance and Training Scheme, vol. 2, Chem. Syst. Chem., 2019, p. e1900031.

[18] H. Guo, Q. Wang, A. Urban and N. Artrith, Artificial Intelligence-Aided Mapping of the Structure–Composition–Conductivity Relationships of Glass–Ceramic Lithium Thiophosphate Electrolytes, vol. 34, Chem. Mater., 2022, p. 6702.

[19] P. M. Piaggi and M. Parrinello, Entropy based fingerprint for local crystalline order, vol. 147, J. Chem. Phys., 2017, p. 114112.

[20] R. Kam, K. Jun, L. Barroso-Luque, J. Yang, F. Xie and G. Ceder, Crystal Structures and Phase Stability of the $Li_2S-P_2S_5$ System from First Principles, vol. 35, Chem. Mater., 2023, p. 9111.

[21] N. D. Lepley, N. A. Holzwarth and Y. A. Du, Structures, Li + mobilities, and interfacial properties of solid electrolytes $Li_3PS_4$ and $Li_3PO_4$ from first principles, vol. 88, Phys. Rev. B, 2013, p. 104103.

[22] J.-S. Kim, W. D. Jung, S. Choi, J.-W. Son, B.-K. Kim, J.-H. Lee and H. Kim, Thermally Induced S‐Sublattice Transition of $Li_3PS_4$ for Fast Lithium-Ion Conduction, vol. 9, J. Phys. Chem. Lett., 2018, p. 5592.

[23] H. Yamada, K. Ohara, S. Hiroi, A. Sakuda, K. Ikeda, T. Ohkubo, K. Nakada, H. Tsukasaki, H. Nakajima, L. Temleitner, L. Pusztai, S. Ariga, A. Matsuo, J. Ding, T.

Nakano and T. Kimur, Lithium Ion Transport Environment by Molecular Vibrations in Ion-Conducting Glasses, vol. 0, Energy Environ. Mater., 2023, p. e12612.

[24] S. Muy, J. C. Bachman, L. Giordano, H. -H. Chang, D. L. Abernathy, D. Bansal, O. Delaire, S. Hori, R. Kanno, F. Maglia, S. Lupart, P. Lamp and Y. Shao-Horn , Tuning mobility and stability of lithium ion conductors based on lattice dynamics, vol. 11, Energy Environ. Sci., 2018, p. 850.

[25] [Online]. Available: http://www.materialsproject.org.

[26] A. Jain, S. P. Ong, G. Hautier, W. Chen, W. D. Richards, S. Dacek, S. Cholia, D. Gunter, D. Skinner, G. Ceder and K. A. Persson, The Materials Project: A materials genome approach to accelerating materials innovation, vol. 1, APL Mater., 2013, p. 011002.

[27] K. Shimizu, E. F. Arguelles, W. Li, Y. Ando, E. Minamitani and S. Watanabe, Phase stability of Au-Li binary systems studied using neural network potential, vol. 103, Phys. Rev. B, 2021, p. 094112.

[28] J. P. Perdew, K. Burke and M. Ernzerhof, Generalized Gradient Approximation Made Simple, vol. 77, Phys. Rev. Lett., 1997, p. 3865.

[29] P. E. Blöchl, Projector augmented-wave method, vol. 50, Phys. Rev. B, 1994.

[30] H. J. Monkhorst and J. D. Pack, Special points for Brillouin-zone integrations, vol. 13, Phys. Rev. B, 1976, p. 5188.

[31] G. Kresse and J. Furthmüller, Efficiency of ab-initio total energy calculations for metals and semiconductors using a plane-wave basis set, vol. 6, Comput. Mater. Sci., 1996, p. 15.

[32] G. Kresse and J. Furthmüller, Efficient iterative schemes for ab initio total-energy calculations using a plane-wave basis set, vol. 54, Phys. Rev. B, 1996, p. 11169.

[33] S. Plimpton, Fast Parallel Algorithms for Short-Range Molecular Dynamics, vol. 117, J. Comput. Phys., 1995, p. 1.

[34] S. Fredericks, K. Parrish, D. Sayre and Q. Zhu, PyXtal: A Python library for crystal structure generation and symmetry analysis, vol. 261, Comput. Phys. Commun., 2021, p. 107810.

[35] P. Virtanen, R. Gommers, T. E. Oliphant, M. Haberland, T. Reddy, D. Cournapeau, E. Burovski, P. Peterson, W. Weckesser, J. Bright, S. J. van der Walt, M. Brett, J. Wilson, K. J. Millman and Mayoro, SciPy 1.0: fundamental algorithms for scientific computing in Python, vol. 17, Nat. Methods, 2020, p. 261.

[36] S. Bringuier. [Online]. Available: https://doi.org/10.5281/zenodo.5261556..

[37] K. Momma and F. Izumi, VESTA 3 for three-dimensional visualization of crystal,

volumetric and morphology data, vol. 44, J. Appl. Cryst., 2011, p. 1272.

[38] A. Stukowski, Visualization and analysis of atomistic simulation data with OVITO–the Open Visualization Tool, vol. 18, Mater. Sci. Eng., 2010, p. 015012.